\documentclass[11pt]{article}
\usepackage{amsfonts}

\begin{document}

\begin{flushright}
FTUV- 07/1607 \quad IFIC/07-35 \quad  hep-th/0707.2336 \quad July 16, 2007. \\
Shorter version (to appear in Phys.Lett.{\bf B}2007)  September 19, 2007
\end{flushright}
\vspace{3.5cm}

\begin{center}

{\LARGE Spinor moving frame, M0--brane covariant BRST quantization
and intrinsic complexity of the pure spinor approach}

\vskip 1.5cm

{\large Igor A. Bandos}

\vskip 1.5cm

{\it Departamento de F\'{\i}sica Te\'orica, Univ.~de Valencia and IFIC
(CSIC-UVEG), 46100-Burjassot (Valencia), Spain}

 and

{\it Institute for Theoretical Physics, NSC ``Kharkov Institute of Physics
and Technology'',  UA61108, Kharkov, Ukraine}

\end{center}

\def\theequation{\arabic{section}.\arabic{equation}}

\vskip 1.5cm

{ To exhibit the possible origin of the inner complexity  of the Berkovits's pure
spinor approach, we consider the covariant BRST quantization of the D=11 massless
superparticle (M0--brane) in its spinor moving frame or twistor-like Lorentz harmonics
formulation. The presence of additional twistor-like variables (spinor harmonics)
allows us to separate covariantly the first and the second class constraints. After
taking into account the second class constraints by means of Dirac brackets and after
further reducing the first class constraints algebra, the dynamical system is described
by the cohomology of a simple BRST charge $\mathbb{Q}^{susy}$ associated to the $d=1$,
$n=16$ supersymmetry algebra. The calculation of the cohomology of this
$\mathbb{Q}^{susy}$ requires a regularization which implies the complexification of the
bosonic ghost associated to the $\kappa$--symmetry and further leads to a complex
(non-Hermitian) BRST charge $\tilde{\mathbb{Q}}^{susy}$  which is essentially the `pure
spinor' BRST charge ${\mathbb{Q}}^{B}$ by Berkovits, but with a composite pure spinor.
}

\thispagestyle{empty}
\newpage

\section{Introduction}

Recently a serious breakthrough in covariant description of quantum superstring theory
has been reached in the framework of the Berkovits  pure spinor approach
\cite{NB-pure}: a technique for loop calculations was developed \cite{NBloops} and the
first results were given in \cite{NBloops,NBloopC}. On the other hand, the pure spinor
superstring was introduced as -and still remains-  a set of prescriptions for quantum
superstring calculations, rather than a quantization of the Green-Schwarz superstring.
In particular, the measure defining the functional intergration over  the pure spinor
ghosts was guessed\footnote{The form of the pure spinor ghost measure appeared in
\cite{NBloops} as a result of  a series of very elegant but indirect arguments
involving the picture changing operator characteristic of the RNS string.} and checked
on consistency \cite{NBloops} rather than derived. Despite a certain progress in
relating the pure spinor superstring \cite{NB-pure} to the original Green--Schwarz
formulation \cite{pure-GS}, and also \cite{Dima+Mario+02} to the superembedding
approach \cite{bpstv,Dima}, the origin and geometrical meaning of the pure spinor
formalism is far from being clear. Possible modifications  of pure spinor formalism are
also considered \cite{GPPPvN,nonmNB}. In particular, an additional non-minimal sector
appeared to be needed to further proceed with loop calculations \cite{nonmNB}. A deeper
understanding of how the pure spinor BRST operator, and other ingredients of the pure
spinor approach,  appear on the way of a straightforward covariant quantization of a
classical action might, in particular, provide a resource of possible non-minimal
variables and give new suggestions in further development of loop calculations.

In this context, the Lorentz harmonic approach
\cite{Sok}--\cite{IB+AN96},
in the frame of which a significant progress toward a covariant superstring
quantization had already been made in late eighties \cite{NPS,Lharm}, looks
particularly interesting. Although no counterpart of the recent progress in loop
calculations \cite{NBloops,NBloopC} has been ever reached in the Lorentz harmonics
framework, its relation with the superembedding approach \cite{bpstv,Dima}, clear
group-theoretical and geometrical meaning \cite{Sok,Wiegmann,B90,Ghsds} and
twistor-likeness \cite{B90,BZ-str,BZ-strH,BZ-p,IB+AN96} suggest it as a promising
starting point of the search for the origin and geometrical meaning of the pure spinor
formalism and its non-minimal modifications. We also hope that the further development
of twistor--like Lorentz harmonic approach, in the pragmatic spirit which characterizes
the pure spinor approach of \cite{NB-pure,NBloops,NBloopC}, might lead to a convenient
and transparent way of the covariant quantum description of superstring. A natural
first stage in such a program is to study the covariant quantization of superparticle,
and in particular, of the {\it $D$=11 massless superparticle}
\cite{B+T=D-branes,Green+99} or {\it M0--brane}, also less studied in comparison with
$D$=10 and $D$=4 superparticle models.

A supertwistor covariant quantization of the massless $D$=11
superparticle has been recently considered in \cite{BdAS2006}. It
starts from twistor-like Lorentz harmonics formulation of the
M0--brane \cite{BL98'}\footnote{See \cite{B90} for $D$=4,
\cite{IB+AN96} for D=10 and \cite{BZ-str,BZ-strH,BZ-p,bst} for the
twistor--like Lorentz harmonic or spinor moving frame formulations
of superstrings, standard and Dirichlet super-p-branes.}, leads to
the linearized $D$=11 supergravity multiplet in the superparticle
quantum state spectrum (in agreement with the light--cone results of
\cite{Green+99}) and exhibits a possible origin of the hidden
$SO(16)$ symmetry of the $D=11$ supergravity \cite{Nicolai87}.

In this letter we report the results of the study of the {\it BRST} quantization of the
$D$=11 massless superparticle in its twistor--like formulation  \cite{BL98',BdAS2006}.
We find a simple reduced BRST charge describing this model and show that the
calculation of its cohomology requires regularization which is made by complexification
of the bosonic ghost for the $\kappa$--symmetry. Then the superparticle spectrum is
described by cohomology of  a {\it complex} BRST charge calculated at vanishing bosonic
ghost. We discuss the relation of this complex BRST charge with the pure spinor BRST
operator by Berkovits. This allows us to explain the intrinsic complexity of the pure
spinor BRST charge. We also present the similar complex Lorentz harmonic BRST charge
for superstring, which is essentially the Berkovits BRST operator but with composite
pure spinors constructed from harmonics and the complexified bosonic ghosts. Derivation
of this BRST operator by covariant quantization of superstring in its spinor moving
frame formulation \cite{BZ-str,BZ-strH} is an interesting problem for future study.

\section{M0-brane in spinor moving frame
formulation}\label{LHSsp}\setcounter{equation}0

The Brink-Schwarz superparticle action can be written in  first order form as
$S_{BS}^{1}= \int_{W^1} (P_{{m}} {\Pi}^{m} -  {1\over 2}d\tau \; e\; P_{{m}} P^{{m}})$.
Here $P_m(\tau)$ is the auxiliary momentum variable, $e(\tau)$ is the worldline einbein
and
\begin{eqnarray}
\label{11DPi} && \Pi^m := dx^m - id\theta \Gamma^m \theta := d\tau
\hat{\Pi}^{m}_\tau \; , \qquad {\Pi}^{m}_\tau:=
\partial_\tau {x}^m(\tau ) - i\partial_\tau {\theta}^\alpha(\tau)
\Gamma^m_{\alpha\beta} {\theta}^\beta (\tau)\;
\end{eqnarray}
is the pull-back of the bosonic supervielbein of flat superspace
(Volkov-Akulov one-form) to the superparticle worldline. The above
formulae are valued in any dimensions. The action of  $D$=11
massless superparticle \cite{B+T=D-branes} is singled out by the
$m=0\, , 1,\ldots 9, \#$ ($\# \equiv 10$) and $\alpha=1, \ldots ,
32$.

The einbein $e(\tau)$ plays the r\^ole of Lagrange multiplier and
produces the mass shell constraint $P_mP^m=0$.  Since this is
algebraic,  if its general solution is known, one may substitute it
for $P_m$ in $S_{BS}^{1}$ and to obtain a classically equivalent
formulation of the $D$- (here 11-) dimensional Brink-Schwarz
superparticle. The moving frame or twistor-like Lorentz harmonics
formulation of \cite{BL98',BdAS2006} (see \cite{B90} for $D$=4 and
\cite{IB+AN96} for  $D$=10) can be obtained just in this way.

It is easy to solve the constraint $P_mP^m=0$ in a non-covariant manner: in a special
Lorentz frame a solution with positive energy reads {\it e.g.} $ P^{\!\!\!^{0}}_{(a)} =
{\rho\over 2} \; (1,\ldots , -1) = {\rho\over 2}  \; (\delta_{(a)}^0
-\delta_{(a)}^{\#})$. The solution in an arbitrary frame follows from this
 by making a Lorentz transformation,
\begin{eqnarray}
\label{PmPm=0->}  P_m := U_m{}^{(a)} P^{\!\!\!^{0}}_{(a)} = {\rho\over 2} (u_{(a)}{}^0
- u_{(a)}{}^{\#}) \, , \qquad U_m{}^{(a)}:= (u_{(a)}{}^0 , u_{(a)}{}^i ,
u_{(a)}{}^{\#}) \in SO(1,D-1) \, . \;\;
\end{eqnarray}
Since  $P_{m}=P_{m}(\tau)$ is dynamical variable in the superparticle action, the same
is true for the Lorentz group matrix $U$ when it is used to express $P_{m}$ through Eq.
(\ref{PmPm=0->}), $U_m{}^{(a)}=U_m{}^{(a)}(\tau)$. Such {\it moving frame variables}
\cite{BZ-str,BZ-strH} are called {\it Lorentz harmonics} \cite{B90,Ghsds} (light--cone
harmonics in \cite{Sok}).

Substituting (\ref{PmPm=0->}) for $P_{m}$ in $S_{BS}^{1}$, one
arrives at the action $ S_{M0}  =  \int_{W^1} \; {1\over 2}
\rho^{++} u^{--}_{{m}} \hat{\Pi}^{m}$ where the vector
$u^{--}_m=u^0_m - u^{\#}_m$ is light--like as follows from the
orthogonality and normalization of the timelike $u_m^{0}$ and
spacelike $u_m^{\# }$ vectors which, in their turn, follow  from
 $U\in SO(1,10)$
in Eq. (\ref{PmPm=0->}).

Moreover, the further analysis shows that the above expression for $S_{M0} $  {\it
hides the twistor--like action}, a higher dimensional (D=11 here) generalization of the
D=4 Ferber--Schirafuji action \cite{Ferber}. Indeed it can be written in the following
equivalent forms \cite{BL98'} (\cite{IB+AN96})
\begin{eqnarray}\label{11DSSP}
S_{M0}&:=& \int d\tau L  =   \int_{W^1} {1\over 2}\rho^{++}\, u_{m}^{--} \, \Pi^m =
\int_{W^1} {1\over 32}\rho^{++}\, v_{\alpha q}^{\; -} v_{\beta q}^{\; -} \, \Pi^m
\tilde{\Gamma}_m^{\alpha\beta}\; ,  \qquad
\\ \nonumber
&& {} \qquad \alpha= 1,2, \ldots , 32 \, \quad (n\; in \; general ) \; , \quad q=1,
\ldots , 16 \, \quad (n/2\; in \; general ) \; ,
\end{eqnarray}
where the first form of the action is as above, while the second
form is twistor--like ({\it cf.} \cite{Ferber}). Instead of
two--component Weyl spinor of the Ferber supertwistor \cite{Ferber},
the action of Eq. (\ref{11DSSP}) includes the set of $16$ bosonic
$32$--component Majorana spinors $v_{\alpha}{}^-_{q}$ which satisfy
the following kinematical constraints (see
\cite{BZ-str,BZ-strH,IB+AN96,BL98'}),
\begin{eqnarray}\label{vv=uG}
\cases{ 2 v_\alpha{}_{q}^{-} v_\beta{}_{q}^{-} =
 u_m^{--}{\Gamma}^m_{\alpha\beta}\quad (a)\;
 , \quad \cr  v_{q}^{-}\tilde{\Gamma}_m v_{p}^{-} =  \delta_{qp} \; u_m^{--}  \quad (b)\;, } \;
v_\alpha{}_{q}^{-}C^{\alpha\beta}v_\beta{}_{p}^{-}=0\quad (c)\;,
   \quad u_m^{--}u^{m --}=0 \qquad (d)\;\; . \qquad
\end{eqnarray}
In \cite{BdAS2006} we presented the supertwistor quantization of the
M0--brane model (\ref{11DSSP}). Here we perform the Hamiltonian
analysis of the system and consider its BRST quantization.

\subsection{Vector and spinor Lorentz harmonics. Spinor moving frame}

Although, in principle, one can study the dynamical system using
just the kinematical constraints (\ref{vv=uG}), it is more
convenient to treat the light--like vector $u_m^{--}$ as an element
of {\it moving frame} and the set of $16$ $SO(1,10)$ spinors
$v_\alpha{}_{q}^{-}$ as part of the corresponding {\it spinor moving
frame}. These moving frame variables are also called ({\it vector}
and {\it spinor}) {\it Lorentz harmonics} (see \cite{GIKOS} for the
notion of harmonics).

The {\it vector} Lorentz harmonics $u_m^{\pm\pm}$, $u_m{}^i$ \cite{Sok} are defined as
elements of the $11\times 11$ Lorentz group  matrix, Eq. (\ref{PmPm=0->}). In the
lightlike basis they are given by
\begin{eqnarray}
\label{harmUin} U_m^{(a)}= (u_m^{--}, u_m^{++}, u_m^{i})\;  \in \; SO(1,10) \; , \qquad
 m= 0,1,\dots,9,\#\; , \quad  i=1,\dots,9 \; ,
\qquad
\end{eqnarray}
where $u^{\pm\pm}_m=u^0_m \pm u^{\#}_m$. The three-blocks splitting (\ref{harmUin}) is
invariant under  $SO(1,1)\otimes SO(9)$; $SO(1,1)$ rotates $u^0_m$ and $u^{\#}_m$ among
themselves and, hence,  transforms their sum and differences, $u^{\pm\pm}_m=u^0_m \pm
u^{\#}_m$, by inverse scaling factors.

The fact that $U\in SO(1,10)$ implies the constraints
\begin{eqnarray}
\label{harmUdef} U^T\eta U = \eta  \quad \Leftrightarrow \cases{ u_m^{--}u^{m--}=0 \; ,
\quad u_m^{++}u^{m++}=0 \; , \quad u_m^{\pm\pm}u^{m\, i}=0 \; , \cr u_m^{--}u^{m++}=2
\; , \qquad u_m^{i}u^{m\, j}=- \delta^{ij} }
\end{eqnarray}
or, equivalently, the unity decomposition
\begin{eqnarray}\label{UUT=eta}
\delta_m^n= {1\over 2}u_m^{++}u^{n--} + {1\over 2}u_m^{--}u^{n++} - u_m^{i}u^{n
i}\qquad  \Leftrightarrow \qquad U\eta U^T=\eta\; .
\end{eqnarray}

The {\it spinor}  harmonics \cite{B90,Ghsds,GHT93} or spinor moving
frame  variables \cite{BZ-str,BZ-strH,BZ-p} $v_{\alpha
q}{}^{\!\!\!\pm}$ are the elements of the $32\times32$ $Spin(1,10)$
matrix
\begin{eqnarray}
\label{harmVin} V_\alpha^{(\beta)}= (v_\alpha{}_q^{-}\; ,
v_\alpha{}_{q}^{+})\; \in \; Spin(1,10)\quad (\alpha=1,\dots 32\; ,
\; q=1,\dots,16) \; .
\end{eqnarray}
They are `square roots' of the associated vector harmonics in the
sense that
\begin{eqnarray}
\label{harmVdef} V \Gamma^{(a)} V^T = \Gamma^m U_m ^{(a)} \; , \qquad V^T
\tilde{\Gamma}_m V = U_m^{(a)} \tilde{\Gamma}_{(a)}\; ,
\end{eqnarray}
which express the $Spin(1,10)$ invariance of the Dirac matrices.

Equation in (\ref{vv=uG}a) is just the $(a)=(--)$ component of the first equation in
(\ref{harmVdef}) taken in the Dirac matrices realization in which $\Gamma^0$ and
$\Gamma^{\# }$ are diagonal and $\Gamma^i$ are off-diagonal. Eq. (\ref{vv=uG}b) comes
from the upper diagonal block of the second equation in Eq. (\ref{harmVdef}). To
complete the set of constraints defining the spinorial harmonics, we have to add the
conditions expressing the invariance of the charge conjugation matrix $C$,
\begin{eqnarray}
\label{harmVdefC}
 VCV^T=C \quad, \quad V^TC^{-1}V=C^{-1}\; ,
\end{eqnarray}
which give rise to the constraint (\ref{vv=uG}c).

In a theory with a local $SO(1,1)\otimes SO(9)$ symmetry containing only one of the two
sets of $16$ constrained spinors (\ref{harmVin}), say $v_{\alpha p}^{\;-}\,$, these can
be treated as homogeneous coordinates of the $SO(1,10)$ coset giving the celestial
sphere $S^9$; specifically (see \cite{Ghsds})
\begin{eqnarray}
\label{v-inS11} {} \{v_{\alpha q}^{\;-}\} = {Spin(1,10) \over [Spin
(1,1)\otimes Spin(9)] \, \subset \!\!\!\!\!\!\times {\mathbb{K}_9} }
= \mathbb{S}^{9}  \quad ,
\end{eqnarray}
where $\mathbb{K}_9$ is the abelian subgroup of $SO(1,10)$ defined
by
 \begin{eqnarray}
\label{K9-def} \delta v_{\alpha q}^{\; -}=0\; , \qquad \delta
v_{\alpha q}^{\; +}=
 k^{++ i} \gamma^i{}_{qp}\,v_{\alpha p}^{\; -}\; , \qquad i=1,\ldots , 9 \; . \qquad
 \end{eqnarray}
Our superparticle model contains just $v_{\alpha q}^{\; -}$ and is invariant under
$SO(1,1)\otimes Spin(9)$ transformations. Hence the harmonics sector of its
configuration space parametrize $S^9$ sphere.

In principle, the constraint Eqs. (\ref{harmUdef}), as equivalent to
 Eq. (\ref{harmUin}), can be solved by
expressing the vector harmonics in terms of $55$ parameters
$l^{(a)(b)}=- l^{(b)(a)}$, $U_m{}^{(a)}=U_m{}^{(a)}(l^{(b)(c)})$,
\begin{eqnarray}
\label{harmU=}  U_m^{(a)}= (u_m^{--}, u_m^{++}, u_m^{i})=\; U_m^{\;(a)}(l^{(c)(d)}) \;
= \delta_m^{(a)} + \eta_{m(b)} l^{(b)(a)} + {\cal O}(l^2)\; . \qquad
\end{eqnarray}
Furthermore, Eqs.  (\ref{harmVdef}), (\ref{harmVdefC}) imply that
spinorial harmonics parametrize the double covering of the
$SO(1,10)$ group element $U_m^{(a)}(l)$ and, hence, that they also
can be expressed through the same $l^{(a)(b)}=- l^{(b)(a)}$
parameters, $V_\alpha^{(\beta)}= V_\alpha^{(\beta)}(l)$,
\begin{eqnarray}
\label{harmV=} V_\alpha^{(\beta)}= (v_\alpha{}_q^{-}\; ,
v_\alpha{}_{q}^{+})\; = \; V_\alpha^{(\beta)}(l^{(a)(b)})\; = \pm
\left(\delta_\alpha^{(\beta)} + {1\over 4}
l^{(a)(b)}\Gamma_{(a)(b)}{} _\alpha^{(\beta)}  + {\cal O}(l^2)
\right)\; . \qquad
\end{eqnarray}

The identification of the harmonics with the coordinates of
$SO(1,10)/H$ corresponds, in this language, to setting to zero the
$H$ coordinates in the explicit expressions (\ref{harmU=}),
(\ref{harmV=}). In our case with $H=[SO(1,1)\otimes SO(9)]\otimes
\mathbb{K}_9$ this implies $l^{0\#}=l^{ij}=l^{++j}=0$ so that the
$SO(1,10)$ matrix is constructed with the use of $9$ parameters
$l^{--j}:= l^{0j}- l^{\# j} $,
\begin{eqnarray}
\label{U=l--} & u_a^{--}= \delta_a^{--} + \delta_a{}^i l^{--i} +
{1\over 2}\delta_a^{++} (l^{--j}l^{--j})\; , \qquad  u_a^{++}=
\delta_a^{++}\; , \qquad u_a{}^{i}= \delta_a{}^{i} + {1\over
2}\delta_a^{++} {l}^{--i} \; . \qquad
\\
\label{V=l--} & v_{\alpha}{}^-_q = \delta_{\alpha}^{-q}  + {1\over
2}\, l^{--i}\gamma^i_{qp} \delta_{\alpha}^{+q} \; , \qquad
v_{\alpha}{}^+_q = \delta_{\alpha}^{+q} \; . \qquad
\end{eqnarray}
In distinction to the general Eqs. (\ref{harmU=}), (\ref{harmV=}), the above equations
are not Lorentz covariant. Although the use of the  explicit expressions
(\ref{harmU=}), (\ref{harmV=})  (their complete form  can be found in
\cite{GomisWest06}) is not practical, it is useful to have in mind the mere fact of
their existence which, in particular, makes transparent that the spinorial and vector
harmonics carries the same degrees of freedom.

\section{M0--brane Hamiltonian mechanics and the BRST charge
$\mathbb{Q}^{susy}$} \setcounter{equation}0

\subsection{Primary constraints of the D=11 massless superparticle model
}\label{Primary}

The phase space $(Z^{{\cal N}}, P_{{\cal N}})$ of our superparticle
model includes the coordinates and momenta
\begin{eqnarray}\label{Pdef}
 & {\cal Z}^{{\cal N}} := \left( x^a ,  \theta^\alpha
,  \rho^{++}\, , \,  \matrix{ U_m^{(a)}  \;  or  \;
{V}{}_{\beta}^{(\alpha)} } \right) \, , \quad P_{_{{\cal N}}} =
{\partial L \over
\partial \dot{{\cal Z}}^{{\cal N}}}  := \left(P_a  ,  \pi_\alpha  ,
 P^{(\rho)}_{++} , \, P_{\;\; (a)}^{[u] \; m} \; or \;
P_{\;(\alpha)}^{[v]\; \beta}\right)\, , \qquad
\end{eqnarray}
restricted by the kinematical constraints (\ref{harmUdef}) or
(\ref{harmVdef}), (\ref{harmVdefC}) and also by the following
primary constraints characteristic of the M0-brane in the spinor
moving frame formulation (\ref{11DSSP})
\begin{eqnarray}\label{P-rvv}
& \Phi_a :=   P_a - {1\over 2} \rho^{++} u_a^{--}\approx 0 \qquad \Leftrightarrow
\qquad \Phi\!\!\!/_{\alpha\beta}:= \Phi_a \Gamma^a_{\alpha\beta}=
P\!\!\!/_{\alpha\beta} - \rho^{++} v_\alpha{}_{q}^{-} v_\beta{}_{q}^{-} \approx 0
\qquad \; , \\ \label{df=}
 & d_\alpha :=  \pi_\alpha + i P\!\!\!/_{\alpha\beta}\theta^{\beta}\approx 0 \; ,
\qquad \pi_\alpha:= {\partial L \over
\partial \dot{\theta}^{\alpha}  } \; , \quad  P_m:= {\partial L \over
\partial \dot{x}^{m} } \\ &
\label{Pr=0} P_{++}^{(\rho )} :=   {\partial L \over
\partial \dot{\rho}^{++}  } \approx 0  \; , \qquad
\\ \label{Pharm=0} &
and \qquad   P^{[u]}{}_{(a)}{}^{m}:=   {\partial L \over
\partial \dot{u}_m^{(a)}  } \approx 0 \;  \qquad or \qquad P^{[v]}{}_{(\alpha)}{}^{\beta}:=
  {\partial L \over
\partial \dot{V}{}_{\beta}^{(\alpha)}  } \approx 0 \; ,
 \end{eqnarray}
Here $\approx$ denotes weak equalities \cite{Dirac}, the equalities which may be used
only after all the Poisson brackets are calculated. This latter are defined by ${}[
P_{_{{\cal M}}} \; , \; {\cal Z}^{{\cal N}}\}_{_{PB}} := - \delta_{\!_{\cal
M}}^{\;{\cal N}} $.

Since  the canonical Hamiltonian $d\tau H_0 := d{\cal Z}^{{\cal N}}\; P_{_{{\cal N}}} -
d\tau \, L \;$ of the massless superparticle is zero in the weak sense, $H_0\approx 0$,
its Hamiltonian analysis reduces to the analysis of the constraints. The presence of
the harmonics in the phase space  (\ref{Pdef}) makes possible to split covariantly the
whole set of the constraints on the first and second class ones (which is not possible
in the original Brink-Schwarz formulation).

\subsection{Second class constraints and Dirac brackets}

Keeping in mind that, upon solving the kinematical constraints (\ref{harmUdef}) and
(\ref{harmVdef}), (\ref{harmVdefC}),  the spinorial and vectorial harmonics are
expressed through the same parameter $l^{(a)(b)}$, Eqs. (\ref{harmU=}), (\ref{harmV=}),
we will use the Language of vector harmonics in the analysis of the bosonic second
class constraints and the spinorial harmonics to separate {\it covariantly} the
fermionic first and second class constraints.

It is convenient to begin with separating the set of 121 primary
constraints $P_{(a)}{}^m \approx 0$ (\ref{Pharm=0}) in a set of 55
constraints $\mathbf{d}_{(a)(b)}:= P_{(a)}{}^m U_{m(b)} -
P_{(b)}{}^m U_{m(a)}$ and the $66$ constraints
$\mathbf{K}_{(a)(b)}:= P_{(a)}{}^m U_{m(b)} + P_{(b)}{}^m U_{m(a)}$
(see \cite{BZ-strH}). The 55 constraints $\mathbf{d}_{(a)(b)}$
commute with the kinematical constraints (\ref{harmUdef}), which we
denote by $\mathbf{\Xi}^{(a)(b)} := U_m^{(a)}U^{m(b)} -
\eta^{(a)(b)}\approx 0$, and generate the Lorentz group algebra
\begin{eqnarray}\label{[d,Xi]=0}
\mathbf{d}_{(a)(b)}:= P_{(a)}{}^m U_{m(b)} - P_{(b)}{}^m U_{m(a)}\approx 0\; , \qquad
 {}[\; \mathbf{\Xi}^{(a)(b)} \; , \; \mathbf{d}_{(a^\prime)(b^\prime)}\; ]_{_{PB}} = 0\;  , \qquad
\\ \label{[d,d]=d}
 {}[\mathbf{d}_{(a)(b)} \; , \; \mathbf{d}^{(c)(d)} \; ]_{_{PB}} = - 4\delta_{[(a)
}{}^{[(c)} \mathbf{d}_{(b)]}{}^{(d)]}\; .  \qquad
\end{eqnarray}
In contrast, the $66$ constraints $\mathbf{K}_{(a)(b)}$ are manifestly second class
ones as far as they are conjugate to the (also second class) $66$ kinematical
constraints (\ref{harmUdef}), ${}[\; \mathbf{\Xi}^{(a)(b)} \; , \;
\mathbf{K}_{(a^\prime)(b^\prime)}\; ]_{_{PB}}  \approx  4 \delta^{((a) }{}_{(a^\prime)}
\delta^{(b))} {}_{(b^\prime)} $
\begin{eqnarray}\label{IIccH}
\mathbf{\Xi}^{(a)(b)} := U_m^{(a)}U^{m(b)} - \eta^{(a)(b)}\approx 0\; , \quad
\mathbf{K}_{(a)(b)}:= P_{(a)}{}^m U_{m(b)} + P_{(b)}{}^m U_{m(a)} \approx 0 \; . \qquad
\end{eqnarray}
At this stage we can introduce Dirac brackets \cite{Dirac} allowing to treat the
constraints (\ref{IIccH}) as strong equalities
\begin{eqnarray}
\label{DBharm}
   {} && [\ldots \; , \; \ldots \}_{_{{DB}^h}}  =  [\ldots \; , \; \ldots \}_{_{PB}}
 - \qquad \nonumber \\ && - {1\over 4}
[\; \ldots \; , \; \mathbf{K}_{(a)(b)} \; ]_{_{PB}} [\; \mathbf{\Xi}^{(a)(b)}\; , \;
\ldots \; ]_{_{PB}}  + {1\over 4} [\; \ldots \; , \; \mathbf{\Xi}^{(a)(b)} \; ]_{_{PB}}
[\; \mathbf{K}_{(a)(b)} \; , \; \ldots \; ]_{_{PB}}  \; ,
\end{eqnarray}

The further study shows the presence of the the following {\it fermionic and bosonic
second class constraints}, the latter split in  mutually conjugate pairs
\begin{eqnarray}
\label{IIcl} & d^+_q:= v^{+\alpha}_q d_\alpha \approx 0 \; , \qquad & \qquad \{ d^+_q
\; , \; d^+_p \}_{_{PB}}= - 2i \rho^{++} \delta_{pq}
 \; , \nonumber \\ & u^{a++}\Phi_a \approx 0 \, , \quad   P_{++}^{[\rho]} \approx 0 \, ,
 \qquad & {}\qquad [u^{a++}\Phi_a \; , \; P_{++}^{[\rho]} \}_{_{PB}}= -1
 \; , \nonumber \\ & u^{a i}\Phi_a \approx 0 \, , \qquad \mathbf{d}^{++j} \approx 0 \, ,
 \qquad  & {}\qquad  [u^{ai}\Phi_a \; , \; \mathbf{d}^{++j}   \}_{_{PB}}=
 - \rho^{++}
 \; . \qquad
\end{eqnarray}
Here $\mathbf{d}^{++ j}= \mathbf{d}^{0 j}+ \mathbf{d}^{\# j}$ is one of the element
appearing in the $SO(1,1)\otimes SO(9)$ invariant splitting of the Lorentz $SO(1,10)$
generator $\mathbf{d}_{(a)(b)}$, $\mathbf{d}^{(a)(b)}= (\mathbf{d}^{(0)}\, ,
\mathbf{d}^{\pm\pm j}\, , \mathbf{d}^{ij})$, $\; v^{+\alpha}_q$ is an element of the
inverse spinor moving frame matrix $V^{-1}{}_{(\beta)}^{\; \alpha}= ( v^{+\alpha}_q \;
, \; v^{-\alpha}_q) \in Spin(1,10)$ which obeys $v^{+\alpha}_qv_{\alpha q}^{\; +}= 0$
and $ v^{+\alpha}_qv_{\alpha q}^{\; -}= \delta_{qp}$. In $D$=11 this is expressed
through the original spinor harmonics by
 $ v^{\pm \alpha}_q = \pm i C^{\alpha\beta}v_{\beta q}^{\;
\pm}$, which is an equivalent form of Eqs. (\ref{harmVdefC}).

Following Dirac \cite{Dirac}, we would like to introduce the Dirac brackets allowing to
treat the second class constraints as strong equalities. For our M0--brane model it is
convenient to do this in two stages (starred and doubly starred brackets in
\cite{Dirac}). On the first stage one introduces the Dirac brackets for sector of
harmonic variables, {\it i.e.} for the second class constraints (\ref{IIccH}),
\begin{eqnarray}
\label{DBharm}
   {} && [\ldots \; , \; \ldots \}_{_{{DB}^h}}  =  [\ldots \; , \; \ldots \}_{_{PB}}
 - \qquad \nonumber \\ && - {1\over 4}
[\; \ldots \; , \; \mathbf{K}_{(a)(b)} \; ]_{_{PB}} [\; \mathbf{\Xi}^{(a)(b)}\; , \;
\ldots \; ]_{_{PB}}  + {1\over 4} [\; \ldots \; , \; \mathbf{\Xi}^{(a)(b)} \; ]_{_{PB}}
[\; \mathbf{K}_{(a)(b)} \; , \; \ldots \; ]_{_{PB}}  \; ,
\end{eqnarray}
while on the second stage one finds the Dirac brackets for all the second class
constraints,
\begin{eqnarray}
\label{DB}
   {} && [\ldots \; , \; \ldots \}_{_{DB}}  =  [\ldots \; , \; \ldots \}_{_{{DB}^h}}
 + \qquad  \nonumber \\ && + [\ldots
\; , \; P_{++}^{[\rho]} ]_{_{PB}} \cdot [ (u^{++}P-\rho^{++}) \; , \; \ldots
]_{_{{DB}^h}}
 - [\ldots \; , \; (u^{++}P-\rho^{++}) ]_{_{{DB}^h}} \cdot [ P_{++}^{[\rho]} \; , \; \ldots
]_{_{PB}} - \qquad \nonumber \\ &&
 - [\ldots \; , \; u^{j}P ]_{_{{DB}^h}} {1\over \rho^{++}} [ \mathbf{d}^{++j}\; , \; \ldots
]_{_{{DB}^h}}
 + [\ldots\; , \; \mathbf{d}^{++j} ]_{_{{DB}^h}} {1\over \rho^{++}} [ u^{j}P\; , \; \ldots
]_{_{{DB}^h}} - \qquad \nonumber \\ &&
 - [\ldots \; , \; d^+_q \}_{_{{DB}^h}} {i\over 2\rho^{++}} [ d^+_q\; , \; \ldots
\}_{_{{DB}^h}} \; .
\end{eqnarray}
Using these Dirac brackets one can treat all the second class constraints as the strong
equalities,
\begin{eqnarray}\label{IIccH=0}
\mathbf{\Xi}^{(a)(b)} := U_m^{(a)}U^{m(b)} - \eta^{(a)(b)}= 0\; , \qquad
\mathbf{K}_{(a)(b)}:= P_{(a)}{}^m U_{m(b)} + P_{(b)}{}^m U_{m(a)} = 0 \; ;  \qquad
\\
\label{strongIIcl}  d^+_q:= v^{+\alpha}_q d_\alpha = 0 \; ; \qquad \rho^{++}=
u^{a++}P_a  \, , \quad   P_{++}^{[\rho]} = 0 \, ;
 \qquad  u^{a i}P_a = 0 \, , \quad \mathbf{d}^{++j} = 0
 \; . \qquad
\end{eqnarray}

\subsection{First class constraints and their algebra} \label{secIclass}

The remaining constraints of the M0--brane model (\ref{11DSSP}),
($\mathbf{d}^{(a)(b)}= (\mathbf{d}^{(0)}\, , \mathbf{d}^{\pm\pm j}\,
, \mathbf{d}^{ij})$, $\mathbf{d}^{(0)}:={1\over 2} \mathbf{d}^{0
\#}$)
\begin{eqnarray}
\label{pre-Icl} & d^-_q:= v^{-\alpha}_q d_\alpha \approx 0 \; ,
\qquad  u^{a--}\Phi_a = u^{a--}P_a =: P^{--} \approx 0 \, ,  \\
\label{pre-IclH} & \mathbf{d}^{ij} \approx 0 \, ,
 \qquad \mathbf{d}^{(0)} \approx 0  \, ,
 \qquad \mathbf{d}^{--i} \approx 0 \; , \qquad
\end{eqnarray}
give rise to the first class constraints. Their Dirac bracket algebra is characterized
by
\begin{eqnarray}
\label{(IH,IH)=DB} {} & [\mathbf{d}^{ij}\; , \;
\mathbf{d}^{kl}]_{_{DB}} = 4\mathbf{d}^{[k|[i} \delta^{j]|l]} \; , &
 [\mathbf{d}^{ij}\; , \; \mathbf{d}^{--k}]_{_{DB}} =
2\mathbf{d}^{-- [i} \delta^{j]k}\; , \quad [\mathbf{d}^{(0)}\; , \;
\mathbf{d}^{\pm\pm i}\}_{_{DB}} = \pm 2 \mathbf{d}^{\pm\pm i} \; , \quad  \\
\label{d--d--DB} && {} \fbox{$[\mathbf{d}^{--i} \; , \; \mathbf{d}^{--j} ]_{_{DB}}  =
  {i\over 2P^{\!^{++}}} \;  d^-_q \gamma^{ij}_{qp}  d^-_p $} \; ,  \qquad \\
\label{(IH,I)=DB} {} & [\mathbf{d}^{ij} \; , \; d^-_p]_{_{DB}} = - {1\over 2}
\gamma^{ij}_{pq} d_q^-  \; , & \qquad [\mathbf{d}^{(0)} \; , \; d^-_p]_{_{DB}} = -
 d_q^-  \; ,  \qquad [\mathbf{d}^{(0)} \; , \; P^{--}]_{_{DB}} = -2
 P^{--}
  \; ,  \qquad
\\
\label{(I,I)=DB} && {} \fbox{$ \; \{ d_q^- \; , \; d^-_{p} \} _{_{DB}}= - 2i
\delta_{qp} P^{--} \;$} \;  .  \qquad {}
\end{eqnarray}
The $16$ fermionic and $1$ bosonic first class constraints in (\ref{pre-Icl}) describe
the {\it irreducible $\kappa$--symmetry}, $d^-_q:=v^{-\alpha}_qd_\alpha $, and its
superpartner ($b$-symmetry), $P^{--}$; these generate the $d=1$, $N=16$ supersymmetry
algebra (\ref{(I,I)=DB}). The irreducibility of the $\kappa$--symmetry in the spinor
moving frame formulation (in contrast with the standard one \cite{ALS}) is due to the
presence of the spinorial harmonics (see \cite{BZ-str,IB+AN96}). The remaining first
class constraints (\ref{pre-IclH}) are originally related to the generators of $[
SO(1,1)\otimes SO(9)]\subset\!\!\!\!\!\!\times K_9]$ subgroup of the Lorentz group
$SO(1,10)$ (see (\ref{[d,d]=d}) with $^{(a)(b)}\not= ^{++ \, i} ,$ $^{i\, ++}$).
However, when passing to Dirac brackets, the deformation in its $[K_9 , K_9]$ part
appears: Eq. (\ref{d--d--DB}) acquires the nonvanishing {\it r.h.s.} proportional to
the product of two fermionic first class constraints (which implies moving outside the
Lie algebra, to the enveloping algebra) \footnote{ This is actually a counterpart of
the well known phenomenon of the non--commutativity of the bosonic coordinate of the
d=4 superparticle which appears in standard formulation of \cite{Casalbuoni} (see also
\cite{JdA+L88}). The appearance of a nonlinear algebra of constraints was also observed
for the $D$=4 null--superstring and null--supermembrane cases \cite{BZnull}. }

One may guess that the complete BRST charge $\mathbb{Q}$  for the algebra of the first
class constraints (\ref{(I,I)=DB}) is quite complicated and its use is not too
practical. Following the pragmatic spirit of the pure spinor approach \cite{NB-pure} we
might take care of the generators of $[SO(1,1)\otimes SO(9)]$ symmetry by imposing them
as conditions on the wavefunctions in quantum theory, calculate the BRST charge
$\mathbb{Q}^\prime$ corresponding to the subalgebra (\ref{d--d--DB}), (\ref{(I,I)=DB})
of $\kappa$--, $b$-- and the deformed $K_9$--symmetry generators, $d_q^-$, $P^{--}$ and
$\mathbf{d}^{--i}$, and study its cohomology  on the space of such wavefunctions.

\subsection{BRST charge
for a nonlinear (sub)algebra, $\mathbb{Q}^{\prime}$, and its
reduction to $\mathbb{Q}^{susy}$} \label{BRST-min} \label{Qprime}

The BRST charge $\mathbb{Q}^{\prime}$ of the nonlinear sub(super)algebra
(\ref{d--d--DB}), (\ref{(I,I)=DB}) of the nonlinear superalgebra of the M0--brane first
class constraints must solve the master equations
\begin{eqnarray}
   \label{QmQm=0}
    {}  \{ \; \mathbb{Q}^{\prime}\; ,  \; \mathbb{Q}^{\prime}\;\}_{_{DB}} =0  \;
\end{eqnarray}
with 'initial conditions' $ \mathbb{Q}^{\prime}\vert_{P^{-[\lambda]}_p= 0\, , \,
\pi^{[c]}_{++}=0\, , \, \pi^{[c]}_{++ j}=0 } = {\lambda}^+_q d_q^{-} + c^{++} P^{--} +
 c^{++j}\mathbf{d}^{--j}$, where
 ${\lambda}^+_q$ is the bosonic ghost for the fermionic
$\kappa$--symmetry, $c^{++}$ and $c^{++j}$ are the fermionic ghosts for the bosonic
$b$--symmetry and deformed $\mathbb{K}_9$ symmetry transformations, and
$P^{-[\lambda]}_q$, $\pi^{[c]}_{++}$ and $\pi^{[c]}_{++j}$ are the (bosonic and
fermionic) ghost momenta conjugate to ${\lambda}^+_q$, $c^{++}$  and $c^{++j}$,
respectively: $\; [{\lambda}^+_q  ,  P^{-[\lambda]}_p ]_{_{DB}} = \delta_{qp}\;$, ${}
\{ c^{++}  , \pi^{[c]}_{++} \} _{_{DB}}= -1\,$, ${} \{ c^{++i} , \pi^{[c]}_{++ j} \}
_{_{DB}}= - \delta^i_{j}\;$. The straightforward calculations show that
$\mathbb{Q}^{\prime}$ does not contain the ghost momentum $\pi^{[c]}_{++j}$  and can be
presented as a sum
\begin{eqnarray}\label{Q'=}
\mathbb{Q}^{\prime}&=& \mathbb{Q}^{susy} + c^{++j}\widetilde{\mathbf{d}}{}^{--j}\;
\end{eqnarray}
of the much simpler BRST charge
\begin{eqnarray}\label{Qbrst1}
\fbox{$\; \mathbb{Q}^{susy}= \lambda^+_q  \; d_q^{-} + c^{++} \;
P^{--} \; - \; i
 \lambda^+_q\lambda^+_q \pi^{[c]}_{++}\;$}\;  , \qquad
{} \{ \mathbb{Q}^{susy} \; , \; \mathbb{Q}^{susy} \}_{_{DB}} = 0 \;
,
 \qquad
\end{eqnarray}
and of the product $c^{++j}\widetilde{\mathbf{d}}{}^{--j}$ of the $c^{++j}$ ghost
fields and the deformed $K_9$ generator modified by additional ghost contributions,
\begin{eqnarray}\label{td--j:=}
\widetilde{\mathbf{d}}{}^{--i} & = {\mathbf{d}}{}^{--i}
 + {i\over 2P^{\!^{++}}} c^{++j}
 d_q^{-}\gamma^{ij}_{qp} P^{-[\lambda]}_p   +
 {1\over P^{\!^{++}}} c^{++j} \lambda_q^{+}\gamma^{ij}_{qp} P^{-[\lambda]}_p
\pi^{[c]}_{++}
 - \nonumber \\ & - {i\over 4(P^{\!^{++}})^2} c^{++j}c^{++k}c^{++l}P^{-[\lambda]}_q\gamma^{ijkl}_{qp} P^{-[\lambda]}_p
  \pi^{[c]}_{++} \; .
   \qquad
\end{eqnarray}

The BRST charge $\mathbb{Q}^{susy} $ (\ref{Qbrst1}) corresponds to the $d=1$, $N=16$
supersymmetry algebra
\begin{eqnarray}\label{16+1al} & {} \{ d_q^{-} \; , \; d_p^-
\}_{_{DB}} = -2i P^{--} \; , \qquad [ P^{--} \; , \; d_p^- ]_{_{DB}}
= 0 \; , \qquad [ P^{--} \; , \; P^{--} ]_{_{DB}} \equiv 0 \; .
 \qquad
\end{eqnarray}
of  the $\kappa$-- and $b$--symmetry generators (\ref{16+1al}). Its `nilpotency' (${}
\{ \mathbb{Q}^{susy},  \mathbb{Q}^{susy} \}_{_{DB}} = 0$) guaranties the consistency of
the reduction of the $\mathbb{Q}^{\prime}$--cohomology problem to the
$\mathbb{Q}^{susy}$--cohomology. As such a reduction is very much in the pragmatic
spirit of the  pure spinor approach \cite{NB-pure,NBloops}, we are going to use it in
this letter \footnote{In classical theory such a reduction can appear as a result of
the gauge fixing, {\it e.g.}, in the explicit parametrization (\ref{harmU=}),
(\ref{harmV=}), by setting $l^{++i}=0=l^{ij}=l^{(0)}$, and  expressing all the
harmonics in terms of $l^{--i}$ by  (\ref{U=l--}), (\ref{V=l--}). } and to study the
cohomology of (\ref{Qbrst1}).

\section{Cohomology of $\mathbb{Q}^{susy}$
and  non-Hermitean  $\tilde{\mathbb{Q}}^{susy}$
charge}\label{QsusyCoH}
 \setcounter{equation}0

\subsection{Quantum M0--brane BRST charge $\mathbb{Q}^{susy}$ and its cohomology problem}

It is practical, omitting the overall  $\pm i$ factor, to write the quantum BRST charge
(\ref{Qbrst1}) as
\begin{eqnarray}\label{Qsusy}
\mathbb{Q}^{susy}= \lambda^+_q  \; D_q^{-} + i c^{++} \partial_{++} -
 \lambda^+_q\lambda^+_q {\partial\over \partial c^{++}}\; , \qquad
{} \{ \mathbb{Q}^{susy} \; , \; \mathbb{Q}^{susy} \} = 0 \; , \qquad
 \qquad
\end{eqnarray}
where the quantum operators $D^-_q$ and $\partial_{++}$, associated
with $d^-_q$ and $P_{++}$, obey ({\it cf.} (\ref{(I,I)=DB}))
\begin{eqnarray}\label{DD=d--}
 {} \{ D_p^{-} , D_q^{-}  \} = 2i \delta_{qp} \partial_{++} \; , \qquad [
 \partial_{++}\, , D_p^{-}]=0 \; ,
\end{eqnarray}
which can be identified with $d=1, n=16$ supersymmetry algebra (or
with its dual which is given by  the algebra of the flat superspace
covariant derivatives). It is convenient to use a realization of
$\partial_{++}$, $D_q^{-}$ as differential operators on the d=$1$,
$n$=$16$ superspace $W^{(1|16)}$ of coordinates
$(x^{++},\theta^+_q)$,
\begin{eqnarray}\label{D-q=}
 D_q^{-} = \partial_{+q}  + i \theta^+_q  \partial_{++}\; , \qquad \partial_{++} :=
 {\partial \over \partial x^{++}}, \qquad \partial_{+q} :=
 {\partial \over \partial \theta^+_q} \; .
\end{eqnarray}
These variables have straightforward counterparts in the so--called
covariant light cone basis, $\theta^+_q= \theta^\alpha
v_{\alpha}{}^+_q$ and $x^{++}=x^{m} u_{m}^{++}$ (see
\cite{Sok,GHT93}).

The Grassmann odd $c^{++}\;$ variable, $\;c^{++}c^{++}=0$, and the
bosonic variables $\lambda^+_q$ in (\ref{Qsusy}) are the ghosts
corresponding to the bosonic and 16 fermionic first class
constraints represented by the differential operators
$\partial_{++}$ and $D^-_q$. Their ghost numbers are $1$, and this
fixes  the ghost number of the BRST charge to be also one,
\begin{eqnarray}\label{ghN}
gh_\# (\lambda^+_q)=1 \; , \qquad gh_\# (c^{++})=1 \; , \qquad  gh_\# (
\mathbb{Q}^{susy})=1\; .
\end{eqnarray}

A non-trivial BRST cohomology is determined by the set of
wavefunctions $\Phi$ of certain ghost numbers $g := gh_\# (\Phi)$
which are BRST-closed, $\mathbb{Q}^{susy}\Phi=0\;$, but not
BRST-exact, $\Phi\not= \mathbb{Q}^{susy}(\ldots )\;$. Moreover, such
functions are defined modulo the BRST transformations {\it i.e.}
modulo BRST-exact wavefunctions $\mathbb{Q}^{susy}\chi$, where
$\chi$ is an arbitrary function of the same configuration space
variables of the ghost number $gh_\# (\chi)= gh_\#(\Phi)-1$ and the
Grassmann parity opposite to the one of $\Phi$,
\begin{eqnarray}
\label{Qcoh=def} \mathbb{Q}^{susy}\Phi=0 \; , \qquad  \Phi \sim
\Phi^\prime = \Phi + \mathbb{Q}^{susy}\chi \; , \qquad gh_\# (\chi)=
gh_\# (\Phi) - 1\; . \qquad
\end{eqnarray}

\subsection{The nontrivial cohomology of $\mathbb{Q}^{susy}$ is located at
$\lambda^+_q\lambda^+_q=0$}
 Decomposing the wave function $\Phi =
\Phi (c^{++} , \lambda^+_q \, ; \, x^{++} , \theta^{+}_q \; , ...)$ in power series of
the Grassmann odd ghost $c^{++}$,
 $\Phi = \Phi_0 + c^{++} \Phi_{++}$,
 one finds that
 $\mathbb{Q}^{susy} \Phi =0$
 for the superfield $\Phi$
 implies
\begin{eqnarray}\label{QPhi=0}
 \lambda^+_q D^-_q \Phi_0 = \lambda^+_q\lambda^+_q \Psi_{++}\quad (a) \; ,
  \qquad \lambda^+_q D^-_q \Psi_{++} = i \partial_{++}\Phi_0 \quad (b)\; .
  \qquad
\end{eqnarray}

Using a similar decomposition for the $\chi$ superfield in
(\ref{Qcoh=def}), $\chi= \chi_0 + c^{++} K_{++}$, one finds
\begin{eqnarray}\label{Phi=Phi+Qchi}
   \Phi \mapsto \Phi^\prime = \Phi + \mathbb{Q}^{susy} \chi \quad \Rightarrow \quad
 \cases{ \Phi_0 \; \mapsto  \Phi_0^\prime = \Phi_0 +  \lambda^+_q D^-_q \chi_0 -
 \lambda^+_q \lambda^+_q  K_{++}\quad (a) \; ,
 \qquad \cr \Psi_{++}  \; \mapsto \Psi_{++}^\prime = \Psi_{++} + i
 \partial_{++}\chi_0 + \lambda^+_q D^-_q K_{++} \quad (b) \;
 }\;  \qquad
\end{eqnarray}
for the BRST transformations.  Using Eqs. (\ref{QPhi=0}), (\ref{Phi=Phi+Qchi}) we can
show that, {\it if one assumes that the spinorial bosonic ghost $\lambda^+_q$ is
non-zero,} or, equivalently, that {\it $\lambda^+_q\lambda^+_q\not= 0$, the BRST
cohomology of $\mathbb{Q}^{susy}$ is necessarily trivial}: all the BRST--closed states
are BRST-exact.

Thus,  if $\mathbb{Q}^{susy}$ has a non-trivial cohomology, it must
have a representation by wavefunctions with support on
$\lambda^+_q\lambda^+_q\not= 0$. In other words, the closed
non-exact wavefunctions representing the non-trivial
$\mathbb{Q}^{susy}$--cohomology must be of the form $\Phi \propto
\delta (\lambda^+_q\lambda^+_q)$ plus a possible
$\mathbb{Q}^{susy}$--trivial contribution.

\subsection{Cohomology at vanishing bosonic ghost and complex BRST operator $\tilde{\mathbb{Q}}^{susy}$}
\label{CohQ-2}

Thus the non-trivial cohomology of $\mathbb{Q}^{susy}$, if exists, must allow a
representation by wavefunctions of the form $\Phi= \delta (\lambda^+_q\lambda^+_q)\;
\Phi^{++}$, where $\Phi^{++}= \Phi^{++}+ c^{++}\Psi^{0}$ has ghost number two units
more than $\Phi$, $\; g_0:= gh_{\#}(\Phi^{++})\, = \, gh_{\#}(\Phi^{0}) + 2\;$. But
there is a difficulty with finding such wavefunctions: since the bosonic ghosts
$\lambda^+_q$ are real, $\lambda^+_q\lambda^+_q=0$ implies $\lambda^+_q=0$. Then, since
${Q}^{susy}$ includes $\lambda^+_q$ in an essential manner, we need in a regularization
allowing us to consider, at the intermediate stages, a nonvanishing $\lambda^+_q$ which
nevertheless obeys $\lambda^+_q\lambda^+_q=0$.

This is possible if we {\it consider $\lambda^+_q$ to be complex} ({\it cf.} with the
pure spinors by Berkovits \cite{NB-pure})
\begin{eqnarray}\label{ll-cl}
\lambda^+_q\mapsto \; \tilde{\lambda}^+_q\; \not= (\tilde{\lambda}^+_q)^*\; \qquad
\Rightarrow \qquad  \tilde{\lambda}^+_q\tilde{\lambda}^+_q=0 \quad \hbox{with }
 \quad \tilde{\lambda}^+_q\not= 0\; \hbox{ is
possible}\;  . \qquad
\end{eqnarray}
{\it The `regularized' BRST charge, $\mathbb{Q}^{susy}_{reg}:=
\mathbb{Q}^{susy}\vert_{\lambda^+\mapsto\tilde{\lambda}^+}$,  is thus non-Hermitian}.
It contains the complex ghost $\tilde{\lambda}^+_q$ rather than the real
${\lambda}^+_q$ in (\ref{Qsusy}), but does not contain $(\tilde{\lambda}^+_q)^*$, and
acts on the space of wavefunctions holomorphic in $\tilde{\lambda}^+_q$. Since the
discussion of the previous section is not affected by above complexification
${\lambda}^+_q\mapsto \tilde{\lambda}^+_q$, we conclude that the non-trivial cohomology
states of the complexified BRST charge can be described by wavefunctions of the form
\begin{eqnarray}\label{Phi(reg)}
 \Phi= \delta
(\tilde{\lambda}^+_q\tilde{\lambda}^+_q)\;
\Phi^{++}(\tilde{\lambda}^+_q \, , \, c^{++}\,  ; \; x^{++} ,
\theta^{+}_q \, , \; \ldots   )\; .
\end{eqnarray}
Now we observe that, as the BRST charge $\mathbb{Q}^{susy}$ does not
contain any derivative with respect to the bosonic ghost
${\lambda}^+_q$, its regularization acts on the $\Phi^{++}$ part of
the function $ \Phi$ in  (\ref{Phi(reg)}) only,
\begin{eqnarray}\label{QsusyPSI=}
\mathbb{Q}^{susy}\vert_{_{{\lambda}^+_p\mapsto \tilde{\lambda}^+_p}}
\;  \delta (\tilde{\lambda}^+_q\tilde{\lambda}^+_q)\;
\Phi^{++}(\tilde{\lambda}^+_q\; \, , \, c^{++}\, ; \ldots ) =
 \delta
(\tilde{\lambda}^+_q\tilde{\lambda}^+_q)\; \tilde{Q}^{susy}
\Phi^{++}(\tilde{\lambda}^+_q\; \, , \, c^{++}\,  ; \ldots ) \; .\qquad
\end{eqnarray}
where we introduced the non-Hermitian  BRST charge
$\tilde{Q}^{susy}=\mathbb{Q}^{susy}\vert_{{\lambda}^+_q\mapsto \tilde{\lambda}^+_q\; :
\; \tilde{\lambda}^+_q\tilde{\lambda}^+_q=0}\;$,
\begin{eqnarray}\label{tQsusy}
\fbox{$\; \tilde{Q}^{susy}= \tilde{\lambda}^+_q  \; D_q^{-} + i c^{++}
\partial_{++} \qquad \; , \qquad
 \tilde{\lambda}^+_q \tilde{\lambda}^+_q  = 0 \; $} \; ,  \qquad
\end{eqnarray}
which is nilpotent, $(\tilde{\mathbb{Q}}^{susy})^2=0$, and can be used to reformulate
the regularized cohomology problem. Note that, once we have  concluded that the
cohomology of $\mathbb{Q}^{susy}$  can be described by wavefunctions of the form
(\ref{Phi(reg)}), we can reduce the nontrivial cohomology search to the set of such
functions, restricting as well the arbitrary superfields $\chi$ of the BRST
transformations (\ref{Phi=Phi+Qchi}) to have the form $\chi = \delta
(\tilde{\lambda}^+_q \tilde{\lambda}^+_q) \chi^{++}$.

Then  the regularized cohomology problem for the complexified BRST operator
($\mathbb{Q}^{susy}$ of (\ref{Qsusy}) now depending on the complexified bosonic ghost
$\tilde{\lambda}^+_q$), reduces to the search for a {\it $\tilde{\lambda}^+_q=0$
`value'} of the  cohomology of the operator $\tilde{Q}^{susy}$ in Eq. (\ref{tQsusy}),
\begin{eqnarray}\label{CHtQsusy}
\tilde{Q}^{susy} \Phi^{++} =0 \; , \qquad \Phi^{++} \sim \Phi^{++\,\prime }= \Phi^{++}
+ \tilde{Q}^{susy}\chi^{++}\;  . \qquad
\end{eqnarray}

This problem (\ref{CHtQsusy}) can be reformulated in terms of
components $\Phi_0^{++}$ and $\Psi^{(0)}$ of the wavefunction
superfield  $\Phi^{++} = \Phi_0^{++} + c^{++} \Psi^{(0)}$ giving
rise to the following equations
\begin{eqnarray}\label{EqCHtQ}
 & \tilde{\lambda}^+_q D^-_q \Phi^{++}_0 = 0\; , \qquad
  \qquad & \tilde{\lambda}^+_q D^-_q \Psi^{(0)} = i \partial_{++}\Phi^{++}_0\; .
  \qquad
\\ \label{Phi=Phi+tQchi}
  & \Phi^{++}_0  \sim  \Phi_0^{++}{}^\prime = \Phi^{++}_0 +
 \tilde{\lambda}^+_q D^-_q \chi^{++}_0 \; ,
 \qquad & \Psi^{(0)}   \sim \Psi^{(0)\prime} = \Psi^{(0)} + i
 \partial_{++}\chi^{++}_0 + \tilde{\lambda}^+_q D^-_q K^{(0)} \; . \qquad
\end{eqnarray}
{\it To obtain the cohomology of $\mathbb{Q}^{susy}$, we have to set
$\tilde{\lambda}^+_q=0$ at the end} to remove the regularization; thus we are really
interested in the wavefunctions for $\tilde{\lambda}^+_q=0$: \\ ${} \quad{} \quad
 \Phi_0^{++}\vert_{\tilde{\lambda}^+_q=0}= \Phi_0^{++}(0 \; , \; x^{++}
, \theta^{+}_q \, ;\; \ldots  )\;$, ${}\quad \Psi_0^{(0)}\vert_{\tilde{\lambda}^+_q=0}=
\Psi_0^{(0)}(0 \; , \; x^{++} , \theta^{+}_q \, ;\; \ldots  )$.

The further study shows that nontrivial `superfield' cohomology problem of Eq.
(\ref{CHtQsusy}) can appear only due to non-triviality of the a
 (pure-spinor like) cohomology problem for the
leading component $\Phi_0^{++}$  of the  $\Phi^{++}$ superfield (see Eqs.
(\ref{EqCHtQ}), (\ref{Phi=Phi+tQchi})),
\begin{eqnarray}\label{CHtQ0}
 \tilde{\lambda}^+_q D^-_q \Phi^{++}_0 = 0\; ,
  \qquad \Phi^{++}_0 & \mapsto  \Phi_0^{++}{}^\prime = \Phi^{++}_0 +
 \tilde{\lambda}^+_q D^-_q \chi^{++}_0 \; . \qquad
\end{eqnarray}
Moreover, we have found that, in its turn, the non-triviality of the
reduced  BRST cohomology (\ref{CHtQ0}) ($(\tilde{\lambda}^+_q
D^-_q$)--cohomology) requires the vanishing ghost number of the
wavefunction  $\Phi^{++}_0$ ($g_0:= gh_{\#} \Phi^{++}_0 =0$) and, in
this case, is described by the kernel $D^-_q\Phi_0^{++}=0$ of the
$\kappa$--symmetry generator,
\begin{eqnarray}\label{gh=0(coh)}
  & g_0:= gh_{\#} \Phi^{++}_0 =0 \; , \qquad   \tilde{\lambda}^+_q D^-_q \Phi^{++}_0 =0   \quad &
  \Rightarrow   \quad D^-_q \Phi^{++}_0 =0 \; .  \qquad
\end{eqnarray}
With the realization (\ref{D-q=}), one finds that the general
solution of this equation is a function independent on both
$\theta^+_q$ and $x^{++}$,
\begin{eqnarray}\label{(gh0coh)=}
 g_0:= gh_{\#} \Phi^{++}_0 =0 \; , \quad
& \Phi^{++}_0 \not= \Phi^{++}_0(x^{++}\, , \, \theta^+_q)   \; \quad
\left({\partial\;\;\; \over \partial x^{++}}\Phi^{++}_0=0\; , \;\;
{\partial\;\;\over
\partial \theta^+_q }\Phi^{++}_0=0\; \right)\;  . \qquad
\end{eqnarray}

Thus the nontrivial {\it cohomology of the BRST charge $\mathbb{Q}^{susy}$}
(\ref{Qsusy}) is described by the cohomology of $\tilde{Q}^{susy}$ (\ref{tQsusy}) in
the sector with (vanishing bosonic ghost and) vanishing ghost number
$g_0:=gh_{\#}(\Phi^{++})=0$ (or $g:=gh_{\#}(\Phi)=-2$ for $\Phi$ in (\ref{Phi(reg)})),
which in
 turn  {\it is  described by the wavefunctions dependent on
the `physical variables' only}. This actually reduces the problem to the quantization
of the physical degrees of freedom, {\it i.e.} to a counterpart of the twistor
quantization of \cite{BdAS2006} which shows that the quantum state spectrum is
described by  the linearized D=11 supergravity multiplet.

\section{Relation with the Berkovits pure spinor BRST charge }
 \label{CohQ-3}

Thus we have shown that the BRST quantization of the M0--brane in
the spinor moving frame formulation (\ref{11DSSP}) leads to the
cohomology problem for {\it complex} BRST charge (\ref{tQsusy}) (the
cohomology at vanishing bosonic ghost gives the M0--brane quantum
state spectrum). Now we turn to the question of relation of our
complex $\widetilde{\mathbb{Q}}^{susy}$, Eq. (\ref{tQsusy}), with
also complex pure spinor BRST operator by Berkovits. This latter
BRST charge has the form \cite{NB-pure}
\begin{eqnarray}\label{QbrstB}\label{NB-pureSp}
\mathbb{Q}^{B}= {\Lambda}^\alpha \; d_\alpha  \; ,
 \qquad  {\Lambda}\Gamma_a{\Lambda}=0 \; , \qquad
    {\Lambda}^\alpha\not= ({\Lambda}^\alpha)^*\; ,
\end{eqnarray}
where $ d_\alpha$ is the fermionic constraint of Eq. (\ref{df=}) and
${\Lambda}^\alpha$ is the complex {\it pure spinor} satisfying the
constraints ${\Lambda}\Gamma_a{\Lambda}=0 $ which guaranties the
nilpotency $(\mathbb{Q}^{B})^2 =0$ of the BRST charge
$(\mathbb{Q}^{B})$.

The $D=11$ pure spinor ${\Lambda}^\alpha$ in general carries $46$ ($23$ complex)
degrees of freedom. A specific $39$ parametric solution $\tilde{\Lambda}$ can be found
using spinorial harmonics $v_{\alpha}{}^-_q$, Eq. (\ref{v-inS11}). It is given by
\begin{eqnarray}\label{Lpure=lv}
 \tilde{\Lambda}_\alpha = \tilde{\lambda}^+_q v_{\alpha}{}^-_q\; \; , \qquad
\tilde{\lambda}^+_q\tilde{\lambda}^+_q=0  \quad  \quad \Rightarrow \quad
\tilde{\Lambda}\Gamma_a\tilde{\Lambda}=0 \; .
 \qquad
\end{eqnarray}
Indeed, as harmonics obey $v^-_q\Gamma_av_p^-= \delta_{qp} u_a^{--}$, Eq.
(\ref{vv=uG}a),
$\tilde{\Lambda}\Gamma_a\tilde{\Lambda}=\tilde{\lambda}^+_q\tilde{\lambda}^+_q$ which
vanishes due to the condition $\tilde{\lambda}^+_q \tilde{\lambda}^+_q = 0$ imposed on
 the complex $16$ component $SO(9)$ spinor $\tilde{\lambda}^+_q$. This latter may be
 identified with the complex zero norm spinor entering the complex charge
$\widetilde{\mathbb{Q}}^{susy}$, Eq. (\ref{tQsusy}).

Furthermore, as far as the $\kappa$--symmetry generator $D^-_q$ is
basically $v^{-\alpha}_q {d}_\alpha$, one finds that our complex
$\widetilde{\mathbb{Q}}^{susy}$ of Eq. (\ref{tQsusy}) is essentially
(up to the simple $c^{++}$ term)  just the Berkovits BRST operator
(\ref{QbrstB}), but with a particular pure spinor
$\tilde{\Lambda}^\alpha$ (\ref{Lpure=lv}) instead of  a generic pure
spinor ${\Lambda}^\alpha$,
\begin{eqnarray}\label{tQsusy=QB}
 \tilde{Q}^{susy} =
 \mathbb{Q}^{B}\vert_{\Lambda^\alpha = \tilde{\lambda}^+_q v^{-\alpha}_q }
 +
i c^{++}
\partial_{++}\; , \qquad
\end{eqnarray}
Thus a counterpart (\ref{tQsusy=QB}) of the Berkovits BRST charge (\ref{QbrstB})
appears when calculating the cohomologies of the regularized version of the BRST charge
(\ref{Qsusy}) which is obtained directly by quantizing the $D=11$ superparticle in the
framework of its twistor--like Lorentz harmonics formulation (\ref{11DSSP}). In our
BRST operator $\tilde{Q}^{susy}$ (\ref{tQsusy=QB}) the zero norm complexified
$\kappa$--symmetry ghost $\tilde{\lambda}^+_q$  carries $30$ of the $39$ degrees of
freedom of the composite the $D=11$ pure spinor  \cite{NB-pure}. The remaining $9$
degrees of freedom in this pure spinor correspond to the $S^9$ sphere of the
light--like eleven--dimensional momentum modulo its energy, parametrized by the
spinorial harmonics, Eq. (\ref{v-inS11}).

Although one may notice the difference in degrees of freedom ($46$ versus $39$), it is
not obvious that all the degrees of freedom in a pure spinor are equally important in
the case of ($D=11$) superparticle. Moreover,  this {\it mismatch disappears in the
`stringy' $D=10$ case} (see below).

In conclusion, let us stress once more that, of all the cohomologies of the complex
Berkovits--like BRST charge $\tilde{\mathbb{Q}}^{susy}$, only their values at vanishing
bosonic ghost, $\tilde{\lambda}^-_q=0$, describe the cohomologies of the M0--brane BRST
charge $\mathbb{Q}^{susy}$ and, hence, the superparticle spectrum. The
$\tilde{\mathbb{Q}}^{susy}$ cohomologies for $\tilde{\lambda}^-_q\not= 0$
(corresponding to nonzero ghost number of the wavefunctions) are reacher and are
related with the spinorial cohomologies of \cite{SpinCohom02}.

\section{Conclusion and  outlook}\label{Concl}

The main conclusion of our present study of the M0--brane case is that the twistor-like
Lorentz harmonic approach \cite{BZ-str,IB+AN96,BdAS2006}, originated in
\cite{Sok,NPS,Lharm}, is able to produce a simple and practical BRST charge. This makes
interesting the similar investigation of the $D=10$ Green--Schwarz superstring case.
For instance, for the IIB superstring the Berkovits BRST charge \cite{NB-pure} looks
like
\begin{eqnarray}\label{QIIB}
\mathbb{Q}^B_{IIB}=\int \Lambda^{\alpha 1}d^1_{\alpha} + \int
\Lambda^{\alpha 2}d^2_{\alpha}\; , \qquad \Lambda^{\alpha
1}\sigma^a_{\alpha\beta}\Lambda^{\beta 1}= 0= \Lambda^{\alpha
2}\sigma^a_{\alpha\beta}\Lambda^{\beta 2}
 \end{eqnarray}
with two complex pure spinors $\Lambda^{\alpha 1}$ and
$\Lambda^{\alpha 2}$ multiplying respectively the left-- and
right--handed stringy counterparts of the superparticle fermionic
constraints (\ref{df=}). By analogy with our study of M0--brane (see
(\ref{Lpure=lv})), one may expect that the BRST quantization of the
spinor moving frame formulation \cite{BZ-str,BZ-strH} of the
Green--Schwarz superstring would lead, after some reduction and on
the way of regularization of the `honest' ('true') hermitian BRST
charge, to the cohomology problem for the complex charge of the form
(\ref{QIIB}) but with composite pure spinors
\begin{eqnarray}\label{pureSp=}
    \widetilde{\Lambda}^{\alpha 1} = \tilde{\lambda}^+_p v^{-\alpha}_p\; , \qquad
 \widetilde{\Lambda}^{\alpha 2} = \tilde{\lambda}^-_p v^{+\alpha}_p\; ,
\qquad    \tilde{\lambda}^+_p\tilde{\lambda}^+_p=0=
\tilde{\lambda}^-_p\tilde{\lambda}^-_p\; . \qquad
\end{eqnarray}
Here $\tilde{\lambda}^{\pm}_p$ are two complex $8$ component $SO(8)$ spinors and the
stringy harmonics $v^{\mp\alpha}_p$ are the homogeneous coordinates of the non--compact
$16$--dimensional coset
\begin{eqnarray}\label{harmV=IIB}
\{ V_{(\beta)}{}^{\alpha} \} = \{ ( v^{-\alpha}_p \; , \; v^{+\alpha}_p )\}  =
{Spin(1,9) \over SO(1,1)\otimes SO(8) } \; ,
\end{eqnarray}
characteristic for the spinor moving frame formulation of the (super)string
\cite{BZ-str,BZ-strH} and describing the spontaneous breaking of the spacetime Lorentz
symmetry by the string model.

It is important that, in distinction to M0--brane case, the $D=10$ solution
(\ref{pureSp=}) of the pure spinor constraints in (\ref{QIIB}) {\it carries the same
number of degrees of freedom} ($44=2\times 8 + 2 \times 14$) that the pair of Berkovits
pure spinors $\Lambda^{\alpha 1}, \Lambda^{\alpha 2}$ ($22+22$). Hence it provides {\it
the general solution} of the $D=10$ pure spinor constraints in terms of harmonics
(\ref{harmV=IIB}) and two complex $SO(8)$ spinors of zero square so that its
substitution for the generic pure spinor of \cite{NB-pure} should not produce any
anomaly or other problem related to the counting of degrees of freedom.

Further development of the present approach is related to the covariant BRST
quantization of superstring in spinor moving frame formulation \cite{BZ-str,BZ-strH}
and to understanding whether/how a cohomology problem for the complex BRST charge
(\ref{QIIB}) appears on this way.

\medskip

{\it\bf Acknowledgments}: The author is thankful to Paolo Pasti,
Dmitri Sorokin, Mario Tonin and especially to Jos\'e A. de
Azc\'arraga for useful discussions. This work has been partially
supported by research grants from the Ministerio de Educaci\'on y
Ciencia (FIS2005-02761) and EU FEDER funds, the Generalitat
Valenciana, the Ukrainian SFFR (N383), the INTAS (2005-7928) and by
the EU MRTN-CT-2004-005104 network in which the author is associated
with the Valencia University.


\bigskip

{\small
\begin{thebibliography}{99}


\bibitem{NB-pure}
 N.~Berkovits,
 {\it Super-Poincare covariant quantization of the superstring},
  JHEP {\bf 0004}, 018 (2000)
  [hep-th/0001035];
  {\it Towards a covariant quantization of the supermembrane},
  JHEP {\bf 0209}, 051 (2002)
  [hep-th/0201151];

\bibitem{NBloops}
N.~Berkovits,
  {\it Multiloop amplitudes and vanishing theorems using the pure spinor
  formalism for the superstring},
  JHEP {\bf 0409}, 047 (2004) [hep-th/0406055];
 \bibitem{NBloopC}
 N.~Berkovits,
  {\it Super-Poincare covariant two-loop superstring amplitudes},
  JHEP {\bf 0601}, 005 (2006);
  {\it New higher-derivative $R^4$ theorems},
  Phys. Rev. Lett. {\bf 98}, 211601 (2007)
  [hep-th/0609006].



\bibitem{pure-GS}
I.~Oda and M.~Tonin,
  {\it On the Berkovits covariant quantization of GS superstring},
  Phys.\ Lett.\ {\bf B520}, 398 (2001)
  [hep-th/0109051];  \\
  N.~Berkovits and D.~Z.~Marchioro,
  {\it Relating the Green-Schwarz and pure spinor formalisms for the
  superstring},
  JHEP {\bf 0501}, 018 (2005)
 [hep-th/0412198].

\bibitem{Dima+Mario+02}
  M.~Matone, L.~Mazzucato, I.~Oda, D.~Sorokin and M.~Tonin,
  {\it The superembedding origin of the Berkovits pure spinor covariant
  quantization of superstrings},
  Nucl.\ Phys.\ B {\bf 639}, 182 (2002)
  [hep-th/0206104].

\bibitem{GPPPvN}
P.~A.~Grassi, G.~Policastro, M.~Porrati and P.~Van Nieuwenhuizen,
  {\it Covariant quantization of superstrings without pure spinor constraints},
  JHEP {\bf 0210}, 054 (2002);
  [hep-th/0112162];
\\ P.~A.~Grassi, G.~Policastro and P.~van Nieuwenhuizen,
   {\it The quantum superstring as a WZNW model},
  Nucl.\ Phys.\  B {\bf 676}, 43 (2004)
  [hep-th/0307056].

\bibitem{nonmNB}
N.~Berkovits and C.~R.~Mafra,
   {\it Some superstring amplitude computations with the non-minimal pure spinor
   formalism},
  JHEP {\bf 0611}, 079, 2006
  [hep-th/0607187]; \\
I.~Oda and M.~Tonin,
    {\it Y-formalism and $b$ ghost in the non-minimal pure spinor formalism of
  superstrings},
 Nucl. Phys. {\bf B779}, 63-100, 2007 [hep-th/0704.1219].



\bibitem{bpstv}
  I.~A.~Bandos, D.~P.~Sorokin, M.~Tonin, P.~Pasti and D.~V.~Volkov,
  {\it Superstrings and supermembranes in the doubly supersymmetric geometrical
  approach},
  Nucl.\ Phys.\  {\bf B446}, 79 (1995)
  [hep-th/9501113].

\bibitem{Dima}
D.~P.~Sorokin,
  {\it Superbranes and superembeddings},
  Phys.\ Rept.\  {\bf 329}, 1 (2000); and refs. therein.





\bibitem{Sok} E.~Sokatchev, {\it Light cone harmonic cuperspace cnd its
applications},
  Phys.\ Lett.\  {\bf B169}, 209-214 (1986).
  {\it Harmonic superparticle},
  Class.\ Quant.\ Grav.\  {\bf 4}, 237-246 (1987).


\bibitem{NPS}
E.~Nissimov, S.~Pacheva and S.~Solomon, {\it Covariant first and second quantization of
the N=2 D=10 Brink--Schwarz Superparticle},
  Nucl.\ Phys.\  {\bf B296}, 462-492 (1988);
{\it  The relation between operator and path integral covariant quantizations of the
Green-Schwarz superstring},
  Phys.\ Lett.\  {\bf B228}, 181-187 (1989).
\bibitem{Lharm}
 R.~Kallosh and M.~A.~Rakhmanov,
 {\it  Covariant quantization of the Green-Schwarz superstring},
  Phys.\ Lett.\  {\bf B209}, 233-238 (1988);
 {\it  Gauge algebra and quantization of type II superstrings},
 {\bf B211}, 71-75 (1988).

\bibitem{Wiegmann}
 P. B. Wiegmann, {\it Multivalued functionals and
geometrical approach for quantization of relativistic particles and strings}, Nucl.
Phys. {\bf  B323}, 311-329 (1989), {\it Extrinsic geometry of superstrings}, Nucl.
Phys. {\bf B323}, 330-336 (1989).

\bibitem{B90}
I.~A.~Bandos,
  {\it A superparticle in Lorentz-harmonic superspace},
  Sov.\ J.\ Nucl.\ Phys.\  {\bf 51}, 906-914 (1990);
{\it Multivalued action functionals, Lorentz harmonics, and spin},
  JETP Lett.\  {\bf 52}, 205-207 (1990).


\bibitem{Ghsds}
  A.~S.~Galperin, P.~S.~Howe and K.~S.~Stelle,
  {\it The superparticle and the Lorentz
  group},
  Nucl.\ Phys.\  {\bf B368}, 248-280 (1992)
  [hep-th/9201020]; \\
 F.~Delduc, A.~Galperin and E.~Sokatchev,
  {\it Lorentz harmonic (super)fields and (super)particles},
  Nucl.\ Phys.\  B {\bf 368}, 143-171 (1992).


\bibitem{BZ-str}
I. A. Bandos and A. A. Zheltukhin, {\it Green-Schwarz
 superstrings in spinor moving frame formalism},
  Phys.\ Lett.\  {\bf B288}, 77-83 (1992).

\bibitem{BZ-strH}
 I.~A.~Bandos and A.~A.~Zheltukhin,
 {\it D = 10 superstring:
Lagrangian and Hamiltonian mechanics in twistor-like Lorentz harmonic formulation},
Phys.\ Part.\ Nucl.\ {\bf 25} (1994) 453-477 [Preprint IC-92-422, ICTP, Trieste, 1992,
81pp.]

\bibitem{BZ-p}
 I.~A.~Bandos and A.~A.~Zheltukhin,
  {\it Generalization of Newman-Penrose dyads in connection with the action
  integral for supermembranes in an eleven-dimensional space},
  JETP Lett.\  {\bf 55}, 81 (1992);
  {\it N=1 superp-branes in twistor--like Lorentz harmonic
  formulation},
  Class.\ Quant.\ Grav.\  {\bf 12}, 609-626  (1995)
  [hep-th/9405113].

\bibitem{GHT93}
 A.~S.~Galperin, P.~S.~Howe and P.~K.~Townsend,
  {\it Twistor transform for superfields},
  Nucl.\ Phys.\  {\bf B402}, 531 (1993).


\bibitem{FedorukZima}
S.~O.~Fedoruk and V.~G.~Zima,
 {\it Covariant quantization of d = 4 Brink-Schwarz superparticle with Lorentz
  harmonics},
  Theor.\ Math.\ Phys.\  {\bf 102}, 305 (1995)
  [hep-th/9409117].

\bibitem{IB+AN96}
I.~A.~Bandos and A.~Y.~Nurmagambetov,
  {\it Generalized action principle and extrinsic geometry for N = 1
  superparticle},
  Class.\ Quant.\ Grav.\  {\bf 14}, 1597-1621 (1997)
  [hep-th/9610098].

\bibitem{B+T=D-branes}
E.~Bergshoeff and P.~K.~Townsend,
  {\it Super D-branes},
  Nucl.\ Phys.\ B {\bf 490}, 145 (1997)
  [hep-th/9611173].

\bibitem{Green+99}
  M.~B.~Green, M.~Gutperle and H.~H.~Kwon,
  {\it Light-cone quantum mechanics of the eleven-dimensional
  superparticle},
  JHEP {\bf 9908}, 012 (1999)
  [hep-th/9907155].


\bibitem{BdAS2006} I.~A.~Bandos,
J.~A.~de Azc\'arraga and D.~P.~Sorokin,
  {\it On D=11 supertwistors, superparticle quantization and a hidden SO(16)
  symmetry of supergravity},
  Proc. XXII
  Max Born Symposium,
  hep-th/0612252.


\bibitem{BL98'} I.~A.~Bandos and J.~Lukierski,
  {\it New superparticle models outside the HLS supersymmetry
  scheme},
  Lect.\ Notes Phys.\  {\bf 539}, 195 (2000)
  [hep-th/9812074] (see Sec. 4, Eq. (4.25) of that paper).

\bibitem{bst}
I.A. Bandos, D.P. Sorokin and M. Tonin, {\it Generalized action principle and
superfield equations of motion for D=10 D--p--branes}, Nucl.Phys. {\bf B497}, 275-296,
1997 [hep-th/9701127].


\bibitem{Nicolai87}
H.~Nicolai,
  {\it D = 11 Supergravity with local SO(16) invariance},
  Phys.\ Lett.\  {\bf B187}, 316 (1987);
  B.~Drabant, M.~Tox and H.~Nicolai,
  Class.\ Quant.\ Grav.\  {\bf 6}, 255 (1989).



\bibitem{Ferber}
A. Ferber, {\it Supertwistors and conformal supersymmetry}, Nucl.\ Phys.\  {\bf B132},
55-64 (1978); 
  T.~Shirafuji,
  {\it Lagrangian mechanics of massless particles with spin},
  Prog.\ Theor.\ Phys.\  {\bf 70}, 18-35 (1983).


\bibitem{GIKOS}
 A.S.~Galperin, E.A.~Ivanov, V.I.~Ogievetsky and E.~S.~Sokatchev, {\it Harmonic
superspace}, Camb. Univ. Press (UK) 2001, 306 pp. and refs. therein.

\bibitem{Dirac}
P.A.M. Dirac, {\it Lectures on quantum mechanics}, Academic Press, NY (1967).



\bibitem{GomisWest06}
J.~Gomis, K.~Kamimura and P.~West,
  {\it The construction of brane and superbrane actions using non-linear
  realisations},
  Class.\ Quant.\ Grav.\  {\bf 23}, 7369-7382 (2006)
  [hep-th/0607057].

\bibitem{ALS}
J. A. de Azc\'{a}rraga and J. Lukierski, {\it Supersymmetric particles with internal
symmetries and central charges}, {Phys. Lett.} {\bf B113}, 170 (1982); {\it
Supersymmetric particles in $N$=2 superspace: phase space variables and Hamiltonian
dynamics},
{\em Phys. Rev.} {\bf D28}, 1337 (1983); \\
 W. Siegel, {\it Hidden local supersymmetry in the supersymmetric particle action},
{ Phys. Lett.} {\bf B128}, 397 (1983).



\bibitem{BZnull}
I.~A.~Bandos and A.~A.~Zheltukhin, {\it Null super $p$-branes quantum theory in
four-dimensional space-time}, Fortsch.\ Phys.\  {\bf 41}, 619 (1993)
and refs. therein.

\bibitem{Casalbuoni}
R.~Casalbuoni, {\it The classical mechanics for bose-fermi systems},
  Nuovo Cim.\   {\bf A33}, 389 (1976).

\bibitem{JdA+L88}
J.~A.~de Azcarraga and J.~Lukierski, {\it Gupta Bleuler quantization of massive
superparticle models in $D$=6, $D$=8 and $D$=10},
  Phys.\ Rev.\  {\bf D38}, 509-513 (1988).





\bibitem{SpinCohom02}
M.~Cederwall, B.~E.~W.~Nilsson and D.~Tsimpis,
  {\it Spinorial cohomology and maximally supersymmetric theories},
  JHEP {\bf 0202}, 009 (2002)
  [hep-th/0110069]; \\
P.S.~Howe and D. Tsimpis,
  {\it On higher-order corrections in M theory},
  JHEP {\bf 0309}, 038 (2003)
  [hep-th/0305129].




\end{thebibliography}
}
\end{document}